\documentclass[sigconf, nonacm]{acmart}

\usepackage{subcaption}

\newcommand{\civ}[1]{\textcolor{black}{\textsc{#1}}}
\newcommand{\cil}[1]{\textcolor{black}{#1}}
\newcommand{\cdv}[1]{\textcolor{black}{\emph{#1}}}

\newcommand{\lp}{\civ{Label Presence}}
\newcommand{\lt}{\civ{Label Type}}
\newcommand{\pc}{\civ{Political Congruence}}

\newcommand{\ltl}{\cil{labeled}}
\newcommand{\ltu}{\cil{unlabeled}}

\newcommand{\lts}{\cil{source}}
\newcommand{\ltk}{\cil{knowledge}}
\newcommand{\ltp}{\cil{propagation}}
\newcommand{\ltg}{\cil{generic}}
\newcommand{\ltn}{\cil{no}}

\newcommand{\pcc}{\cil{congruent}}
\newcommand{\pci}{\cil{incongruent}}

\newcommand{\lik}{\cdv{Like}}
\newcommand{\com}{\cdv{Comment}}
\newcommand{\sha}{\cdv{Share}}

\AtBeginDocument{
  \providecommand\BibTeX{{
    \normalfont B\kern-0.5em{\scshape i\kern-0.25em b}\kern-0.8em\TeX}}}

\copyrightyear{2023}
\acmYear{2023}
\setcopyright{acmlicensed}\acmConference[HAI '23]{International Conference on Human-Agent Interaction}{December 4--7, 2023}{Gothenburg, Sweden}
\acmBooktitle{International Conference on Human-Agent Interaction (HAI '23), December 4--7, 2023, Gothenburg, Sweden}
\acmPrice{15.00}
\acmDOI{10.1145/3623809.3623856}
\acmISBN{979-8-4007-0824-4/23/12}

\begin{document}

\title[Automated Misinformation Warning Labels on Post Engagement Intents]{Effects of Automated Misinformation Warning Labels on the Intents to Like, Comment and Share Posts}

\author{Gionnieve Lim}
\email{gionnievelim@gmail.com}
\orcid{0000-0002-8399-1633}
\affiliation{
  \institution{Singapore University of Technology and Design}
  \streetaddress{8 Somapah Rd}
  \country{Singapore}
  \postcode{487372}
}

\author{Simon T. Perrault}
\email{perrault.simon@gmail.com}
\orcid{0000-0002-3105-9350}
\affiliation{
  \institution{Singapore University of Technology and Design}
  \streetaddress{8 Somapah Rd}
  \country{Singapore}
  \postcode{487372}
}

\renewcommand{\shortauthors}{Lim and Perrault}

\begin{abstract}
With fact-checking by professionals being difficult to scale on social media, algorithmic techniques have been considered. However, it is uncertain how the public may react to labels by automated fact-checkers. In this study, we investigate the use of automated warning labels derived from misinformation detection literature and investigate their effects on three forms of post engagement. Focusing on political posts, we also consider how partisanship affects engagement. In a two-phases within-subjects experiment with 200 participants, we found that the generic warnings suppressed intents to comment on and share posts, but not on the intent to like them. Furthermore, when different reasons for the labels were provided, their effects on post engagement were inconsistent, suggesting that the reasons could have undesirably motivated engagement instead. Partisanship effects were observed across the labels with higher engagement for politically congruent posts. We discuss the implications on the design and use of automated warning labels.
\end{abstract}

\ccsdesc[100]{Human-centered computing~Empirical studies in HCI}

\keywords{misinformation, automated fact-checking, warning labels, partisanship, social media engagement}

\maketitle

\section{Introduction}

Misinformation is prevalent on social media platforms. Capitalizing on the global networks that enable quick and extensive spread of information, motivated actors have taken to using the platforms for purposes like earning money through clickbaits or influencing public opinion through disinformation campaigns, often spreading sensationalist and false content~\cite{Mourao2019}. To address misinformation on social media platforms, a countermeasure is engaging fact-checking professionals to verify and apply labels on problematic content~\cite{Mosseri2016}. Many studies have looked into the effectiveness of these labels, often focusing on how they affect the perceived accuracy of posts and the intent to share them~\cite{Rathje2021, Oeldorf-Hirsch2023}. While this countermeasure has been found to be generally effective, it is difficult to scale~\cite{Moy2021}, and other methods such as crowdsourced and algorithmic fact-checking have been proposed~\cite{Zhou2020}.

Studies have looked into the potential of using automated fact-checking to label misinformation. In a study on two types of labels on Twitter~\cite{Lanius2021}, one indicating that the user was a suspected bot account and another indicating that the tweet contained misinformation, both were observed to decrease participants' willingness to engage with tweets and to change some participants' perceptions toward the misinformation. In a study comparing warnings from different sources~\cite{Seo2019}, the machine-learning warning which disputed the post was found to perform worse than the fact-checking warning by humans. But when the machine-learning warning was extended to include a graph that displayed the factors attributing to the algorithm's decision, participants were observed to discern between fake and true news most accurately. Another study had corresponding observations where both algorithmic and third-party fact-checker labels were found to reduce participants' perceived accuracy and believability of fake posts irrespective of the post’s political ideology~\cite{Jia2022}.

Our study differs and extends these studies in two ways. First, we investigate a different set of warning labels that are derived from the main categories of algorithmic misinformation detection in the literature. Second, we investigate the effects of three common forms of post engagement, including the intents to comment on and like posts, apart from just sharing them. In doing so, our study offers a closer look at how engagement patterns vary across the engagement intents and how the reasons for the labels can affect them differently. As political misinformation has been of concern in recent years~\cite{Jerit2020}, we look at political posts, and thereby also consider partisanship effects~\cite{Altay2023} on post engagement. Our work supports the literature that automated misinformation labeling is a viable measure for scaling fact-checking but that careful consideration must be put in both the design of the labels as well as in ensuring that users have a basic understanding of the underlying algorithm so as not to misinterpret or be misled by the warning labels.

In this paper, we report the findings of our study that involves a two-phases within-subjects experiment with 200 participants. We look at the effects of automated misinformation warning labels on the intents to like, comment on and share posts, and also consider how these effects changed when different reasons were provided for the labels. With partisanship being a key factor of engaging with political posts, we also consider how the effects of the labels were affected by the political congruence of the posts to the participants. In the remainder of the paper, we discuss the literature grounding our work, the results of the two experiment phases, and their implications, thereby contributing to literature informing the use of automating fact-checking for labeling content on social media.

\section{Related Work}

This work is informed by literature on misinformation labeling interventions, social media engagement and partisanship.

\subsection{Misinformation Labeling Interventions}

On social media platforms, one countermeasure of misinformation is the use of fact-checking labels that are applied to problematic content by professional human fact-checkers. As fact-checking is an effort intensive process that involves many steps and factors to manually verify the content~\cite{Uscinski2013}, this countermeasure is challenged by being unable to scale well~\cite{Moy2021}. To address this, alternatives such as crowdsourcing and algorithmic measures have been considered~\cite{Zhou2020}.

With increased computational power and easy access to large volumes of data, there has been strong interest in developing misinformation detection algorithms using machine learning techniques~\cite{Zhou2020}. These algorithms rely on a variety of features such as the linguistic style of the content, platform metadata like the time, location and creators of posts, and the scope and speed of the spread of the content~\cite{Wu2019, Islam2020}. 

While misinformation detection can reach fairly high levels of accuracy~\cite{Mridha2021}, automated fact-checking remains used largely for downranking or removing content on social media platforms instead of for labeling purposes~\cite{Saltz2021b}. This is despite the fact that studies have found labels by automated fact-checkers to be effective in reducing the perceived accuracy of and engagement with false content~\cite{Seo2019, Lanius2021, Jia2022}. While a possible reason for this is social media platforms being cautious, we posit that by making clear that the labels only warn about potential misinformation, there is room for automated misinformation labeling using warnings instead of fact-checking labels which would allow for misinformation to be addressed in a timely and scalable manner.

\subsection{Engaging with Social Media Content}

Social media platforms offer a variety of ways for users to interact with content such as being able to like, comment on and share posts. To make profits, a priority of these platforms is to maximize the time that users spend on them~\cite{Martens2018}. One way this is achieved is through the use of recommendation algorithms to show content that users are likely to be interested in on their feeds. To personalize the feed, the algorithm uses posts that users have engaged with to recommend similar content that matches their interests. While recommendation algorithms are effective in bringing relevant content to the users, a drawback is that they can also confine the scope of information and perspectives that the users are exposed to, thereby contributing to the formation of narrow worldviews~\cite{Jamieson2010}.

Another cause for concern with social media engagement lies in how the metrics are interpreted by users. Geeng et al.~\cite{Geeng2020} found that users tended to pay more attention to posts with greater community engagement. Tandoc et al.~\cite{Tandoc2018} further observed that the engagement is viewed as a form of validation by the community that the content is reliable. With post engagement metrics affecting the content that is shown to users and to their networks, and also influencing users' perceptions of the reliability of posts, the impact of labeling interventions on suppressing engagement has become a key measure of their effectiveness.

In evaluating labeling interventions, a large body of work have focused on the intent to share posts due to its direct impact on the spread of misinformation whereas there have been fewer studies investigating the intents to like and comment on posts (e.g., \cite{Simko2019, Geeng2020, t'Serstevens2022}). We thus seek to contribute to the understanding of this lesser explored space by investigating the effect of labels on all three forms of engagement. 

\subsection{Partisanship}

Partisan identity strongly predicts individuals' preferences towards social policy issues in the US~\cite{PRC2014}. The growing political divide has also led to greater contempt towards the opposing party and those who identify with it. On social media, this partisanship has influenced the browsing and post engagement behavior of users. There is a greater preference towards clicking and reading content aligned with one's political ideology especially among hyper-partisan individuals~\cite{Peacock2021}. Studies on sharing behavior have also found that individuals were more likely to do so for content on the opposing party, usually to express negative emotions~\cite{Rathje2021} or to defame them~\cite{Oeldorf-Hirsch2023}. With partisanship being a strong factor affecting the engagement with political content on social media, we further consider the effectiveness of the labels on suppressing engagement with posts that are politically congruent and incongruent to participants' political identities.

\section{Method}

We conducted a two-phases within-subjects experiment. In the first phase, we sought to test how labels provided by automated fact-checkers affected the intents to like, comment on and share posts compared to those that were unlabeled. We expected that there would be lower intent to engage with labeled false posts than with unlabeled false posts. In the second phase, we tested whether having reasons provided for the labels affected the engagement intents and in what ways they did so. We expected that regardless of the reason, having a label would suppress engagement intents for false posts, however, the magnitude of suppression would differ based on the reason provided by the automated fact-checker. For both phases, we further explored the interaction of the labels with the political congruence of the posts to the users since previous literature has shown that partisanship affects post engagement~\cite{Rathje2021, Oeldorf-Hirsch2023}. The research questions were thus:

\begin{enumerate}
    \item How does the presence of automated warning labels affect post engagement?
    \item How do different reasons for the automated warning labels affect post engagement?
    \item How do automated warning labels affect post engagement when partisanship is taken into account?
\end{enumerate}

\subsection{Procedure}

Participants accessed the study using a web app developed by the researchers. They progressed through the following stages: consent to participate, demographics survey, first experiment phase and second experiment phase. The study was approved by the University's Institutional Review Board (IRB).

\begin{figure}[!htb]
    \centering
    \includegraphics[width=.95\linewidth]{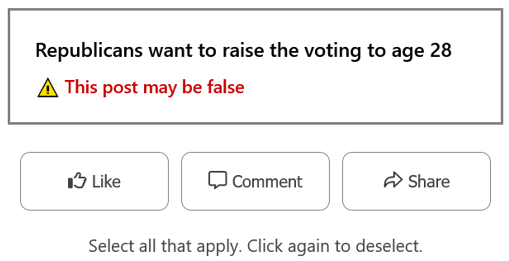}
    \caption{The experiment interface showing a post with the \ltg\ label and the engagement options to \lik, \com\ and \sha.}
    \label{fig:maininterface}
\end{figure}

In the first phase of the experiment aimed at comparing the presence and absence of automated warning labels, participants were told that \textit{``A computer program that automatically checks and detects suspicious posts has been developed. Imagine that you are browsing through your social media feed where the automated fact-checker is being used to check the posts.''} They were then presented with four posts (two labeled and two unlabeled false posts) and could choose to like, comment and share each post (Figure~\ref{fig:maininterface}). A label that simply hints towards a problematic post without providing any reason for it, i.e., \textit{``This post may be false''}, is used.

In the second phase of the experiment that aimed to compare the automated warning labels when the reasons for them were provided, participants were further told that \textit{``The automated fact-checker now gives the top reason for why it thinks the post may be false.''} They then continued to rate another twelve posts (6 labeled false posts and 6 unlabeled true posts). The labeled posts would have any one of the three labels (Knowledge, Source, Propagation) shown in Table~\ref{tab:labels}. The inclusion of unlabeled true posts (which are not used for any analysis) was meant to balance the number of labeled to unlabeled posts to provide a more realistic social media browsing experience.

\begin{table}[!htb]
  \caption{Types of warning labels. The `Reference' column indicates how each label is referred to in the paper.}
  \label{tab:labels}
  \begin{tabular}{lp{.7\linewidth}}
    \toprule
     Reference & Label\\
    \midrule
    Generic & This post may be false\\
    Knowledge & A significant number of posts with content similar to this are reported as false\\
    Source & A significant number of other posts made by this account are reported as false\\
    Propagation & A significant number of shares of this post is spreading it widely very quickly\\
    \bottomrule
    \end{tabular}
\end{table}

In both experiments, the orders of the presented posts and labels, and which false post a label was applied to were fully randomized for counterbalancing. Each participant would be exposed to two instances of each label in the posts they read.
The posts were shown in an ``infinite scroll'' layout which, along with the fixed engagement buttons, were meant to replicate the feeds of social media platforms like Facebook and Twitter.

\subsection{Materials}

\paragraph{News}
A set of 32 news headlines was collected where each subset of eight belonged to a combination of political lean \{pro-Democrat/pro-Republican\} $\times$ news veracity \{true/false\}. All false and some true news were gathered from fact-checking websites (e.g., PolitiFact, Snopes) that had verified and rated the veracity of the news. As there was a lack of fact-checked true news, we also looked at mainstream news websites (e.g., ABC News, CNN) to gather more of them. We assumed that the news would be true as the same content was reported across multiple sites. Most news was published within a month of the study and no longer than three months. Where necessary, we made minor modifications to the news for a consistent reporting style.

\paragraph{Labels}
There are four warning labels (Table~\ref{tab:labels}). The \ltg\ label simply warns about the post and the phrase ``may be'' is used to indicate uncertainty since automated fact-checking does not have the level of certainty of professional human fact-checking. The \ltk, \lts\ and \ltp\ labels provide reasons for warning about the post. These labels were chosen by consulting survey papers on algorithmic misinformation detection~\cite{Zhou2020, Ansar2021}. Together, the papers offered five categories of detection strategies (Table~\ref{tab:detectionmethods}) but we excluded Stance and Style as these covered information that a user would have been able to see directly from a social media post.

\begin{table}[!htb]
  \caption{Strategies used to detect misinformation in machine learning algorithms.}
  \label{tab:detectionmethods}
  \begin{tabular}{lp{.7\linewidth}}
    \toprule
    Method & Description\\
    \midrule
    Knowledge & Check the content of the post compared to external information from other sources\\
    Source & Check the creator of the post and the users spreading it\\
    Propagation & Check the spread of the post (e.g., how fast and widely it spreads)\\
    Stance & Check the opinions of users engaging with the post (e.g., through the likes, shares and comments)\\
    Style & Check the tone and writing style of the post\\
    \bottomrule
\end{tabular}
\end{table}

\begin{figure}[!htb]
    \centering
    \includegraphics[width=.95\linewidth]{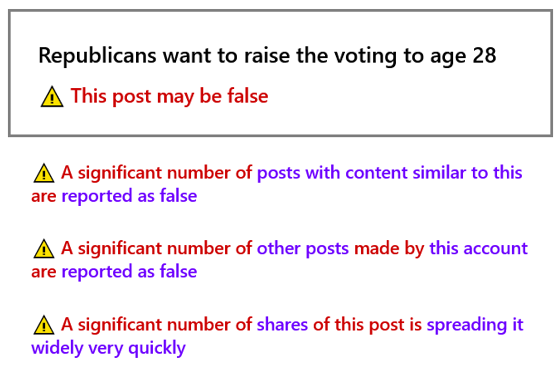}
    \caption{An example post with the \ltg\ label. Examples of the \ltk, \lts\ and \ltp\ labels are shown below.}
    \label{fig:mainnewslabels}
\end{figure}

\paragraph{Posts}
Each post contains a piece of news with either a label or no label (Figure~\ref{fig:mainnewslabels}). The news is in black text while the label is prefixed with a warning emoji and is in smaller red and purple text. Purple is used to highlight the differences between labels as a pilot study found that using only red text led users to neglect the differences in the content. All texts are wrapped in a fixed size box with a black border.

\subsection{Recruitment and Participants}

Participants were recruited on Amazon Mechanical Turk, a crowd work platform. They had to be at least 18 years old, based in the US, with an approval rate of 99\% and at least 10,000 approved human intelligence tasks. A compensation of USD 7.25 per hour was given following our IRB guidelines.

The study involved 200 participants\footnote{Twenty participants were excluded due to poor quality responses such as failing the attention checks.}. Their average age was 43.3 ($SD=11.9, Min=24, Max=75$), and a majority identified as male (56.5\%) and White (86.5\%). Most participants had a bachelor's degree (51.0\%). During the recruitment process, we screened for participants that used social media via a multiple-selection question with the options for mainstream social media platforms. We also screened them according to their self-reported political affiliation, reported via a single-selection question with the options: Democrat, Independent, Republican, to reach an equal number of Democrat to Republican participants\footnote{Participants that identified as Independent were not recruited.}.

The sample size of 200 was determined through power analyses set to detect a small effect size (d=.02) at 80\% power. We expected a small effect size informed by the findings of experimental studies that assessed similar forms of misinformation interventions~\cite{Clayton2020, Lees2022}.

\section{Results}

There are two parts to our analysis\footnote{The results for true posts were not used for the analysis.}. The first phase (Phase I) compared labeled and unlabeled false posts. This comparison will allow us to determine whether the \lp\ (two levels: \ltg\ label and \ltn\ label) has an effect on users' post engagement.

The second phase (Phase II) compared different types of automated warning labels on false posts. We refer to this as \lt\ (four levels: \ltg, \ltk, \lts\ and \ltp\ labels). The results of the \ltg\ label from the first phase is used as the control for the second phase. We excluded the unlabeled posts in the first phrase from the analysis so as to focus only on labeled content.

We also looked at the interaction effects of \lp\ and \lt\ with the \pc\ of the news to the participants (two levels: \pcc\ and \pci) respectively. If the news is pro-Democrat and the participant identifies as Democrat, the news is categorized as \pcc\ for the participant and so forth.

We measured participants' intents to \lik, \com\ and \sha\ posts using a 2-point scale \{0: No, 1: Yes\} based on whether the respective engagement buttons were selected. There were thus three dependent variables for post engagement.

As our data did not follow a normal distribution, Aligned Rank Transform (ART) tests~\cite{Wobbrock2011} was used for the non-parametric analysis of the main effects and the interactions and we report the $F$-statistic, $p$-value and effect size ($\eta^2_p$). Contrast tests~\cite{Elkin2021} with Benjamini-Hochberg corrections to minimize false discovery rates~\cite{Glickman2014} were used for the analysis of pairwise comparisons and are reported with the $t$-statistic. Means ($M$) and standard deviations ($SD$) of the dependent variables are reported as well.

\subsection{Phase I: How Does The Presence of Labels Affect Engagement Intents?}

The first experiment tested \lp\ and the interaction with \pc\ on the intents to \lik, \com\ and \sha\ posts. The results of the ART tests are in Table~\ref{tab:H1ART}. The intent to comment was found to be significantly lower for \ltl\ (\ltg\ label) than \ltu\ (\ltn\ label) posts. Having labels also seemed to dampen the intent to share, but this was not observed for the intent to like.

\begin{table}[!htb]
  \caption{Effects of the presence of warning labels on the engagement intents. Scores for the intent to like/comment/share posts are between 0 and 1.}
  \label{tab:H1ART}
  \begin{tabular}{cccc}
    \toprule
    Engagement & $F(1, 597)$ & $p$-value & $\eta^2_p$\\
    \midrule
    Like & 1.37 & .24 & .0023\\
    Comment & 4.64 & .032* & .0077\\
    Share & 3.14 & .077 & .0052\\
  \bottomrule
\end{tabular}

\vspace{.5cm}
  \begin{tabular}{ccc}
    \toprule
    & Generic & No label\\
    \cline{2-3}
    Engagement & $M (SD)$ & $M (SD)$\\
    \midrule
    Like & .080 (.27) & .075 (.26)\\
    Comment & .20 (.40) & .25 (.43)\\
    Share & .025 (.16) & .038 (.19)\\
  \bottomrule
\end{tabular}
\end{table}

\subsection{Phase II: How Do Different Reasons for the Labels Affect Engagement Intents?}

The second experiment tested providing different reasons for the labels (\lt) and the interaction with \pc\ on the intents to \lik, \com\ and \sha\ posts. The results of the ART tests are in Table~\ref{tab:H2ART}. A significant main effect for the intent to share was observed. Contrast tests for the intent to \sha\ show that the \ltp\ label was significantly higher than all the other label types: \ltg\ ($t_{1393}=-4.44, p<.001$), \ltk\ ($t_{1393}=-3.61, p=.0010$) and \lts\ ($t_{1393}=-3.26, p=0023$).

\begin{table}[!htb]
  \caption{Effects of different warning labels on the engagement intents. Scores for the intent to like/comment/share posts are between 0 and 1. The superscripts ($^{\alpha}, ^{\beta}, ^{\gamma}$) represent significant pairwise differences.}
  \label{tab:H2ART}
  \begin{tabular}{cccc}
    \toprule
    Engagement & $F(3, 245)$ & $p$-value & $\eta^2_p$\\
    \midrule
    Like &  1.96 & .12 & .0042\\
    Comment & 0.30 & .83 & .00064\\
    Share & 7.59 & <.001* & .016\\
  \bottomrule
  \end{tabular}

  \vspace{.5cm}
    \begin{tabular}{ccccc}
    \toprule
    & Generic & Knowledge & Source & Propagation\\
    \cline{2-5}
    Engagement & $M (SD)$ & $M (SD)$ & $M (SD)$ & $M (SD)$\\
    \midrule
    Like & .080 (.27) & .080 (.27) & .062 (.24) & .085 (.28)\\
    Comment & .20 (.40) & .20 (.40) & .21 (.41) & .21 (.40)\\
    Share & .025 (.16)$^{\alpha}$ & .043 (.20)$^{\beta}$ & .045 (.21)$^{\gamma}$ & .075 (.26)$^{\alpha \beta \gamma}$\\
  \bottomrule
  \end{tabular}
\end{table}

An interaction effect with \pc\ was observed for the intent to \com\ ($F(1,1393)=12.13, p<.0001$) although contrast tests did not reveal any significant pairwise comparisons. An interaction effect with \pc\ was also observed for the intent to \sha\ ($F(1,1393)=128.78, p<.0001$) and the results of the contrast tests are in Table~\ref{tab:H2CT}. For the \ltp\ label, there was significantly higher intent to share posts that were \pcc\ rather than \pci.

\begin{table}[!htb]
  \caption{Results of the contrast tests for \pc\ $\times$ \lt\ on the intent to share. Scores for the intent to share posts are between 0 and 1.}
  \label{tab:H2CT}
  \begin{tabular}{ccccc}
    \toprule
    & & & \multicolumn{2}{c}{Political Congruence}\\
    & & & Congruent & Incongruent\\
    \cline{4-5}
    Label Type & t(1393) & p-value & M (SD) & M (SD)\\
    \midrule
    Generic & 0.54 & .82 & .030 (.17) & .020 (.14)\\
    Knowledge & 0.27 & .85 & .045 (.21) & .040 (.20)\\
    Source & 0.54 & .82 & .050 (.22) & .040 (.20)\\
    Propagation & 3.26 & .0053* & .11 (.31) & .045 (.21)\\
  \bottomrule
\end{tabular}
\end{table}

\section{Discussion}

We discuss our findings and their implications on the design and use of automated misinformation warning labels. We first describe the effects of political congruence on engagement and then delve into the effects of the labels themselves.

\subsection{Engagement Intents are Generally Higher for Politically Congruent Posts}

Consistent with the literature, we observed partisanship effects on post engagement. For instance, Table~\ref{tab:H2CT} shows that there is a higher intent to share congruent posts across all label types.
The observed partisan bias in engaging with posts aligns with work on other forms of misinformation interventions. Gao et al.~\cite{Gao2018} found that stance labels intensified selective exposure whereby participants were more likely to click and read news politically aligned with themselves. Oeldorf-Hirsch et al.~\cite{Oeldorf-Hirsch2020} found that participants were more likely to share negative content about opposing parties even with the presence of fact-checking labels. Partisanship could explain these observations where actions are motivated by ``feeling rather than thinking''~\cite{Peacock2021}. The effects of partisanship may even overshadow the efforts of misinformation interventions. Grady et al.~\cite{Grady2021} found that providing pre-warnings for false political headlines was effective in discouraging belief even for politically congruent items, however, participants regressed to their initial beliefs in a follow-up study two weeks later when presented with the same headlines without the warnings. With political news and misinformation being generated at a pace faster than human fact-checking can possibly catch up with, it is important to not just quickly, but also consistently address them, and automated interventions can be a measure to meet these needs. It is also important to raise here that we discuss this in light of the US political scene and Western context and that more work would be needed to understand if our observations hold in the political scenes of other countries with different cultural contexts. An interesting follow up could be to conduct this study in South Korea, an Eastern country where strong partisan divide is also observed~\cite{Silver2022}.

\subsection{Engagement Intents Are Affected By Automated Warning Labels In Different Ways} 

From Table~\ref{tab:H1ART}, we observed that the generic warning label dampened the intent to comment. The trend is the same even when the knowledge, source and propagation labels are presented instead (all three have similar values compared to the generic label in Table~\ref{tab:H2ART}). On the other hand, while the generic label somewhat dampened the intent to share, the knowledge, source and propagation labels seemed to encourage sharing instead. While these two intents exhibit different behaviors, we posit that a similar rationale could have driven them. As commenting is an effortful task, users who would not want to waste time engaging with potentially false content may avoid commenting~\cite{Kim2017}. However, as sharing is a low effort task, users would simply have to press a few buttons for purposes like cautioning their networks about the false content~\cite{Tandoc2020}. Interestingly, there was an inconsistent effect of the labels on the intent to like. Only the \lts\ label seemed to suppress the intent to like which was not observed in the other labels. The source of the content has been shown to be a strong indicator of its reliability~\cite{Heuer2022} which may have explained this. Overall, the findings suggest that there may be underlying differences in the motivations to like, comment on and share posts. This is further exemplified by how the intents to comment were generally higher and consistent across all the labels compared to the intents to like and share (Table~\ref{tab:H2ART}). Perhaps when it came to commenting, participants cared more about the content of the post, whereas for liking and sharing, the labels were taken into greater account. Qualitative work to understand the reasons for engaging with labeled posts, particularly whether the motivation lies with the post or the label (e.g., sharing to show support of the original content vs sharing to inform their network that the labeled content is false) is a line of investigation that is lacking in existing literature, and would be necessary to better inform and validate our conjectures.

\subsection{Label Descriptions Are Held In Higher Regard Despite A Common Fact-Checking Algorithm}

Participants were informed that the same automated fact-checker was used throughout the study and that the label showed the top reason for the decision. Other reasons would also have impacted the automated fact-checker's warning, just to a lesser extent. While all posts would have been run through the same automated fact-checker, participants exhibited different behavior based largely on the reason shown for the warning. This is most evident in the intent to share where there was significantly higher intent to share posts with the \ltp\ label compared to all the other labels (Table~\ref{tab:H2ART}).
Given that unlike the \ltg, \ltk\ and \lts\ labels that hint more strongly that a post is suspicious, the reason by the \ltp\ label is vaguer as rapid spreading could also indicate breaking news or virality among other cases. As such, participants could have been less skeptical of the post and even felt encouraged to join others in sharing the content~\cite{Borges-Tiago2019}. This observation raises several considerations in having an automated fact-checker explain its decision. First, it is important to inform or educate users about the underlying fact-checking algorithm so that they understand that several features impact the decision and not just a sole feature. Second, the wording of the labels has to be carefully crafted as participants are likely to interpret them as is which may lead to undesired effects instead.

\section{Limitations}

As our study is conducted in an experimental setting, there are limitations on its ecological validity. We sought to ameliorate this by designing the engagement buttons and rating interface based on popular social media platforms to provide a more realistic browsing experience. Another limitation is whether the intent to engage with posts would translate to actual engagement. While a study on sharing intent has found such a pattern~\cite{Mosleh2020}, it would also be important to verify this for the intents to like and comment.

\section{Conclusion}
There has been rising interest in investigating a variety of labeling interventions for misinformation in social media platforms. While these platforms currently engage human fact-checkers to label posts, the issue of scaling this process has been well noted and automated fact-checking has been proposed as a viable alternative. In this study, we look at how automated misinformation warning labels and the reasons provided by the automated fact-checker for the label affect people's engagement with posts. While related studies have largely focused on the intent to share posts among other forms of engagement, we further consider how warning labels affect the intent to like and comment on them.

Conducting a two phase within-subjects experiment with 200 participants, we found that the presence of warning labels suppressed the intents to comment on and share posts, but not to like them. When the labels are accompanied with the reasons for them, we found participants to be influenced by the reasons even though they were informed that the same algorithm was used to power the decision of applying the label. Depending on the reason, the label was less effective, particularly the \ltp\ label. We also observed interaction effects of partisanship with there being higher intent to share politically congruent posts.

Our results surface several considerations regarding the evaluation of misinformation interventions and the design of automated warning labels. Apart from just the intent to share, researchers interested in such studies should also consider the effectiveness of the labeling interventions on suppressing the intents to like and comment as they can be influenced differently. Care should be taken when designing the reasons shown to participants in the warning labels and people should have a basic understanding of the underlying algorithm so as not to misinterpret or be misled by the warnings. Where political congruence effects may strongly sway beliefs and with political news constantly being created, automated labeling interventions can be valuable in promptly addressing accompanying misinformation.

\bibliographystyle{ACM-Reference-Format}
\bibliography{main}

\end{document}